\documentclass[12pt]{article}
\catcode`\@=11

\@addtoreset{equation}{section}
\def\theequation{\arabic{section}.\arabic{equation}}
     
\catcode`\@=11
\def\thesection{\arabic{section}}

\def\appendix{\setcounter{section}{0}
        \def\thesection{Appendix.}
        \def\theequation{\Alph{section}.\arabic{equation}}}
\def\section{\@startsection{section}{1}{\z@}{3.5ex plus 1ex minus
   .2ex}{2.3ex plus .2ex}{\large\bf}}
     
\def\eqnarray{\let\@currentlabel=\theequation\refstepcounter{equation}
    \global\@eqnswtrue
    \global\@eqcnt\z@\tabskip\@centering\let\\=\@eqncr
    $$\halign to \displaywidth\bgroup\@eqnsel\hskip\@centering
      $\displaystyle\tabskip\z@{##}$&\global\@eqcnt\@ne 
       \hfil${{}##{}}$\hfil
      &\global\@eqcnt\tw@ $\displaystyle\tabskip\z@{##}$\hfil 
       \tabskip\@centering&\llap{##}\tabskip\z@\cr}
\def\lefteqn#1{\hbox to 4\arraycolsep{$\displaystyle #1$\hss}}
\long\def\@makefntext#1{\parindent 0cm\noindent
\hbox to 1em{\hss$^{\@thefnmark}$}#1}

\def\IR{{\hbox{{\rm I}\kern-.2em\hbox{\rm R}}}}
\def\IH{{\hbox{{\rm I}\kern-.2em\hbox{\rm H}}}}
\def\IC{{\ \hbox{{\rm I}\kern-.6em\hbox{\bf C}}}}
\def\IZ{{\hbox{{\rm Z}\kern-.4em\hbox{\rm Z}}}}

\newcommand{\beq}{\begin{equation}}
\newcommand{\eeq}{\end{equation}}

\begin{document}

%%%%%%%%%%%%%%%%%%%%%%%%%%%%%%%%%%%%%%%%%%%
%     C I T E . S T Y
%     compressed lists of numerical citations: [11-16]
%     see also OVERCITE.STY and DRFTCITE.STY
%
%     Copyright (C) 1989-1992 by Donald Arseneau
%     These macros may be freely transmitted, reproduced, or modified for
%     non-commercial purposes provided that this notice is left intact.
%
%
%  \@citen contains the code that parses the list of names, ignoring
%  spaces after commas, writes the aux file \citation, and formats the
%  number list.  \citen can be used by itself to give citation numbers
%  without the other formatting; e.g., "See also ref.~\citen{junk}."
%
\def\citen#1{%
\edef\@tempa{\@ignspaftercomma,#1, \@end, }% ignore spaces in parameter list
\edef\@tempa{\expandafter\@ignendcommas\@tempa\@end}%
\if@filesw \immediate \write \@auxout {\string \citation {\@tempa}}\fi
\@tempcntb\m@ne \let\@h@ld\relax \let\@citea\@empty
\@for \@citeb:=\@tempa\do {\@cmpresscites}%
\@h@ld}
%
% for ignoring spaces in the input:
\def\@ignspaftercomma#1, {\ifx\@end#1\@empty\else
   #1,\expandafter\@ignspaftercomma\fi}
\def\@ignendcommas,#1,\@end{#1}
%
% For each citation, check if it is defined, if it is a number, and
% if it is a consecutive number that can be represented like 3-7.
%
\def\@cmpresscites{%
 \expandafter\let \expandafter\@B@citeB \csname b@\@citeb \endcsname
 \ifx\@B@citeB\relax % undefined
    \@h@ld\@citea\@tempcntb\m@ne{\bf ?}%
    \@warning {Citation `\@citeb ' on page \thepage \space undefined}%
 \else%  defined
    \@tempcnta\@tempcntb \advance\@tempcnta\@ne
    \setbox\z@\hbox\bgroup % check if citation is a number:
    \ifnum\z@<0\@B@citeB \relax
       \egroup \@tempcntb\@B@citeB \relax
       \else \egroup \@tempcntb\m@ne \fi
    \ifnum\@tempcnta=\@tempcntb % Number follows previous--hold on to it
       \ifx\@h@ld\relax % first pair of successives
          \edef \@h@ld{\@citea\@B@citeB}%
       \else % compressible list of successives
%         % use \hbox to avoid easy \exhyphenpenalty breaks
          \edef\@h@ld{\hbox{--}\penalty\@highpenalty \@B@citeB}%
       \fi
    \else   %  non-successor--dump what's held and do this one
       \@h@ld \@citea \@B@citeB \let\@h@ld\relax
 \fi\fi%
 \let\@citea\@citepunct
}
%
%%    To put space after the comma, use:
\def\@citepunct{,\penalty\@highpenalty\hskip.13em plus.1em minus.1em}%
%%    For no space after comma, use:
%% \def\@citepunct{,\penalty\@highpenalty}%
%%
%
%  Make \@citex refer to \citen:
%
\def\@citex[#1]#2{\@cite{\citen{#2}}{#1}}%
%
%  Replacement for \@cite.  Give one normal space before the citation,
%  set high penalties for linebreaks,
%
\def\@cite#1#2{\leavevmode\unskip
  \ifnum\lastpenalty=\z@ \penalty\@highpenalty \fi % highpenalty before
  \ [{\multiply\@highpenalty 3 #1% % triple-highpenalties within list
      \if@tempswa,\penalty\@highpenalty\ #2\fi % and before note.
    }]\spacefactor\@m}
\let\nocitecount\relax  % in case \nocitecount was used for drftcite
%
%%%%%%%%%%%%%%%%%%%%%%%%%%%%%%%%%%%%%%%%%%%% 
\begin{titlepage}
\vspace{.5in}
\begin{flushright}
UCD-99-17\\
gr-qc/9909087\\
September 1999\\
revised December 1999\\
\end{flushright}
\vspace{.5in}
\begin{center}
{\Large\bf
Aberration and the Speed of Gravity}\\
\vspace{.4in}
{S.~C{\sc arlip}\footnote{\it email: carlip@dirac.ucdavis.edu}\\
       {\small\it Department of Physics}\\
       {\small\it University of California}\\
       {\small\it Davis, CA 95616}\\{\small\it USA}}
\end{center}

\vspace{.5in}
\begin{center}
{\large\bf Abstract}
\end{center}
\begin{center}
\begin{minipage}{4.4in}
{\small
The observed absence of gravitational aberration requires that ``Newtonian'' gravity
propagate at a speed $c_g>2\times10^{10}c$.  By  evaluating the gravitational effect 
of an accelerating mass, I show that aberration in general relativity is almost exactly 
canceled by velocity-dependent interactions, permitting $c_g=c$.  This cancellation 
is dictated by conservation laws and the quadrupole nature of gravitational radiation.
}
\end{minipage}
\end{center}
%\begin{flushleft}
%\small PACS: 04.20.-q, 04.25.-g, 03.50.De \\
%\small Keywords: Gravitation; Aberration; Causality; Retardation; Faster-than-light
%\end{flushleft}
\end{titlepage}
\addtocounter{footnote}{-1}

In a recent paper in Physics Letters {\bf A} \cite{TvF}, Van Flandern has argued that
observations show that gravity propagates at a speed much greater than $c$.  In the 
absence of direct measurements of propagation speed, Ref.\ \cite{TvF} relies instead 
on directional information, in the form of observations of (the absence of) gravitational
aberration.  But the translation from a direction to a speed requires theoretical assumptions, 
and the implicit assumptions of Ref.\ \cite{TvF}---in particular, that the interaction is 
purely central, with no velocity-dependent terms---do not hold for general relativity,
or, for that matter, for Maxwell's electrodynamics.  

In this paper, I explicitly compute the gravitational effect of an arbitrarily accelerating 
source, Kinnersley's ``photon rocket'' \cite{Kinn}.  Although gravity propagates at the 
speed of light in general relativity, the expected aberration is almost exactly canceled by 
velocity-dependent terms in the interaction.  While at first this cancellation seems to be 
``miraculous,'' it can be explained from first principles by turning Van Flandern's argument 
on its head: conservation of energy and angular momentum, together with the quadrupole 
nature of gravitational radiation, require that any causal theory have such a cancellation.

\section{Aberration in Electromagnetism \label{s1}}

It is certainly true, although perhaps not widely enough appreciated, that observations 
are incompatible with Newtonian gravity with a light-speed propagation delay added 
in \cite{Good,Light}.  If one begins with a purely central force and puts in a finite
propagation speed by hand, the forces in a two-body system no longer point toward
the center of mass, and the resulting tangential accelerations make orbits drastically 
unstable.  A simple derivation is given in problem 12.4 of Ref.\ \cite{Light}, where it 
is shown that Solar System orbits would shift substantially on a time scale on the order 
of a hundred years.  By analyzing the motion of the Moon, Laplace concluded in 1805
that the speed of (Newtonian) gravity must be at least $7\times10^6 c$ \cite{Laplace}.  
Using modern astronomical observations, Van Flandern raised this limit to 
$2\times10^{10}c$ \cite{TvF}.

But this argument, at least in its simplest form, holds only if one postulates that the 
relevant force is purely central and independent of the source velocity.  As Poincar{\'e} 
observed as early as 1905 \cite{Poin,Poin2}, the effects of aberration can be drastically 
altered by velocity-dependent interactions.  And indeed, for Maxwell's electrodynamics 
and Einstein's general relativity, such interactions occur \cite{Marsh}.

As a warm-up, let us first consider electrodynamics.  It is well known 
that if a charged source moves at a constant velocity, the electric field experienced by a 
test particle points toward the source's ``instantaneous'' position rather than its retarded 
position.  Lorentz invariance demands that this be the case, since one may just as well 
think of the charge as being at rest while the test particle moves.  This effect does not 
mean that the electric field propagates instantaneously; rather, the field of a moving 
charge has a velocity-dependent component that cancels the effect of propagation 
delay to first order \cite{Feynman}.

It is helpful to analyze this case a bit more carefully, while establishing notation that will 
be useful below when we discuss general relativity.  Let the source move along a timelike
world line $C$ in flat Minkowski spacetime, with position $z^\mu(s)$ and four-velocity 
$\lambda^\mu = dz^\mu/ds$.  The backwards light cone from any point $x^\mu$ will 
intersect $C$ at a point $z^\mu(s_R)$ (see figure~1), 
\begin{figure}
\begin{picture}(150,150)(-140,-10)
\qbezier(31,41)(45,80)(32,110)
\qbezier(32,110)(25,130)(29,142)
\qbezier[8](31,41)(29,35)(29,27)
\qbezier(29,26)(30,5)(37,-20)
\put(90,120){\circle*{3}}
\put(94,118){$x$}
\put(90,120){\line(-1,-1){100}}
\put(90,120){\line(1,-1){100}}
\qbezier(-10,20)(90,40)(190,20)
\qbezier(-10,20)(90,0)(190,20)
\put(31,41){\circle*{3}}
\put(35,40){$z(s_R)$}
\put(31,41){\line(3,4){60}}
\put(31,41){\vector(3,4){30}}
\put(32,140){$z(s)$}
\put(62,76){$\sigma^\mu(x)$}
\end{picture}
\caption{The geometry of retarded positions in Minkowski space}
\end{figure}
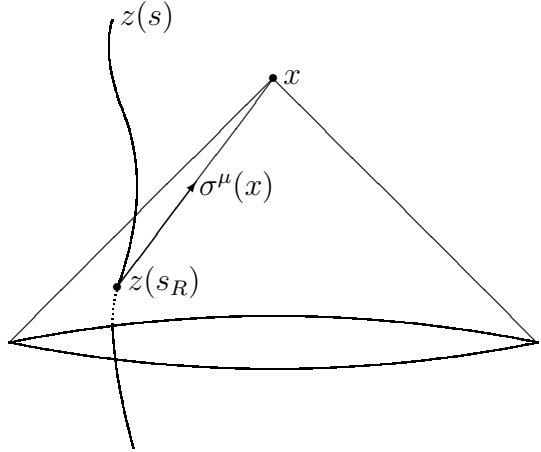
and this relation can be viewed as
an implicit definition of the retarded proper time $s_R(x)$:
\beq
\eta_{\mu\nu} (x^\mu - z^\mu(s_R))(x^\nu - z^\nu(s_R)) = 0 .
\label{a1}
\eeq
Let
\beq
\sigma^\mu = x^\mu - z^\mu(s_R)
\label{a2}
\eeq
denote the null vector connecting $x$ and $C$.  Differentiating (\ref{a1}), we obtain
\beq
\partial_\mu s_R(x) = {\sigma_\mu\over r} ,
\label{a2a}
\eeq
where
\beq
r(x) = \lambda^\mu(s_R)\sigma_\mu 
\label{a3}
\eeq
is an invariant retarded distance from $x$ to $C$.
In terms of a (3+1)-dimensional decomposition of spacetime, we have (in units $c=1$)
\begin{eqnarray}
\sigma^0 = R, \quad & \sigma^i = Rn^i \nonumber\\
\lambda^0 = \gamma_R, \quad &\lambda^i = \gamma v_R^i ,
\label{a4}
\end{eqnarray}
where $R = |{\bf x} - {\bf z}(s_R)| = t - z^0(s_R)$ is the retarded spatial distance,
${\bf v}_R$ is the retarded velocity, $\gamma_R = (1-v_R^2)^{-1/2}$, and ${\bf n}$ 
is a unit spatial vector pointing toward the retarded position of the source.  In 
``propagation-delayed Newtonian gravity,'' aberration appears as the fact that the 
force is directed along $\bf n$, and not along the vector pointing toward the 
``instantaneous'' position of the source.

With these conventions,  the Li{\'e}nard-Wiechert potential in Maxwell's electrodynamics 
can be written as \cite{Jackson}
\beq
A^\mu = {e\over r}\lambda^\mu(s_R) .
\label{a5}
\eeq
Using standard identities \cite{Kinn,Damour} obtained from eqn.\ (\ref{a2a}), one 
obtains a field strength tensor
\beq
F_{\mu\nu} = \partial_\mu A_\nu - \partial_\nu A_\mu
   = {e\over r^3}\left(1-\sigma^\rho{d\lambda_\rho\over ds}\right)
   \left(\sigma_\mu\lambda_\nu - \sigma_\nu\lambda_\mu\right) 
   + {e\over r^2}\left(\sigma_\mu{d\lambda_\nu\over ds} 
   - \sigma_\nu{d\lambda_\mu\over ds}\right) .
\label{a6}
\eeq
In particular, eqn.\ (\ref{a4}) implies that the electric field can be written as
\beq
E^i = F^{i0} = {e\over\gamma_R^2 R^2(1-{\bf n}\cdot{\bf v}_R)^3} (n^i - v^i_R)
   + \hbox{\it radiative terms} ,
\label{a7}
\eeq
where the omitted ``radiative'' terms depend explicitly on acceleration and fall off as
$1/R$ rather than $1/R^2$.  

Note that every term in eqns.\ (\ref{a6})--(\ref{a7}) is retarded, and that nothing 
depends on the ``instantaneous'' position or direction of the source.  The potential 
(\ref{a5}) similarly depends only on purely retarded quantities; while a different 
gauge choice such as Coulomb gauge may lead to an ``instantaneous'' term in the 
potential, this is illusory, since all {\em physical\/} quantities will continue to depend 
only on retarded characteristics of the source \cite{Brill}.  

Nevertheless, the direction of the nonradiative ``Coulomb'' force in (\ref{a7}) is
\beq
n^i - v^i_R = (1-{\bf n}\cdot{\bf v}_R)
   \left(n^i + (t - z^0(s_R)){dn^i\over dt}\right) .
\label{a8}
\eeq
The second term in this expression is essentially a linear extrapolation from the
retarded direction $n^i$ toward the ``instantaneous'' direction.  In particular, for 
a  charge in uniform motion it is easy to check that $n^i-v^i_R$ points toward the 
``instantaneous'' position, so the effects of aberration are exactly canceled.

Does eqn.\ (\ref{a7}) imply that the electric field propagates instantaneously?   
Clearly  not.  In particular, if a uniformly moving charge suddenly stops at
position $z(s_0)$, the field at a distant location $x$ will continue to point toward 
its ``extrapolated'' position---even though the charge never actually reaches that
position---until the time $t-z(s_0)$ that it takes for light to travel from $z(s_0)$
to $x$.  At that time, the field will abruptly switch direction to point toward the true 
position of the source.  This sudden change in the field, propagating outward from
$z(s_0)$ at the speed of light, is what we mean by the electromagnetic radiation of 
an accelerated charge.  (For a simple derivation of electromagnetic radiation as
the retarded effect of the changing Coulomb field of an accelerated charge, see
Appendix B of Ref.\ \cite{Purcell}.)  One could, of course, try to formulate an 
alternative model in which the Coulomb field acted instantaneously, but only at 
the expense of ``deunifying'' Maxwell's equations and breaking the connection 
between electric fields and electromagnetic radiation.

\section{Aberration in Gravity}

If we try to extend the arguments of the preceding section to general relativity, we face 
two subtleties.  First, there is no preferred time-slicing in general relativity, and thus 
no unique definition of an ``instantaneous'' direction.  For weak fields, we can use 
the nearly flat background to define a nearly Minkowski coordinate system, but we 
must expect ambiguities of order $v^2$.  Second, we cannot simply require by fiat that 
a massive source accelerate.  The Einstein field equations are consistent only when 
all gravitational sources move along the trajectories determined by their equations 
of motion, and in particular, we can consistently represent an accelerated source only 
if we include the energy responsible for its acceleration.

Fortunately, an exact solution for such an accelerated source exists.  Kinnersley's
``photon rocket'' \cite{Kinn, Damour, Bonnor} represents a mass with an arbitrary
acceleration brought about by the nonisotropic emission of electromagnetic
radiation.  Its metric, in the notation of the preceding section, is
\beq
g_{\mu\nu} = \eta_{\mu\nu} - {2Gm(s_R)\over r^3}\sigma_\mu\sigma_\nu .
\label{b1}
\eeq
This metric contains four arbitrary functions of time: a time-varying mass $m$ and the 
three independent components of the acceleration $d\lambda^\mu/ds$.  In general, it 
has a nonvanishing stress-energy tensor proportional to $\sigma_\mu\sigma_\nu$, 
representing radiation or null dust streaming out from the world line $C$; it reduces to 
the Schwarzschild metric, with a vanishing stress-energy tensor, when $m$ is constant 
and $C$ is a straight line.

A test particle in the spacetime (\ref{b1}) will travel along a geodesic.  If we use the flat
metric $m=0$ to define background Minkowski coordinates, the ``acceleration'' of such a
particle, in Newtonian language,  is determined by the connection $\Gamma^\rho_{\mu\nu}$.
In particular, if the particle is initially at rest, its ``acceleration'' is  $-\Gamma^i_{00}$.
A long but routine computation yields\footnote{Note that there is a sign error in  Ref.\ 
\cite{Kinn}.}
\begin{eqnarray}
\Gamma^\rho_{\mu\nu} = &-&{2Gm\over r^3}\eta_{\mu\nu}\sigma^\rho + 
   {Gm\over r^4}(3\lambda_\mu\sigma_\nu\sigma^\rho + 3\lambda_\nu\sigma_\mu\sigma^\rho
   -\sigma_\mu\sigma_\nu\lambda^\rho) \nonumber\\
   &-& {3Gm\over r^5}\left(1 - \sigma_\tau{d\lambda^\tau\over ds}\right)
   \sigma_\mu\sigma_\nu\sigma^\rho - {2G^2m^2\over r^6}\sigma_\mu\sigma_\nu\sigma^\rho
   - {1\over r^4}G{dm\over ds}\sigma_\mu\sigma_\nu\sigma^\rho ,
\label{b1a}
\end{eqnarray}
and in particular,
\begin{eqnarray}
\Gamma^i_{00} = {Gm\over R^2}{1\over\gamma_R^3(1-{\bf n}\cdot{\bf v}_R)^5}\biggl[&&
   \left(1 - 2{\bf n}\cdot{\bf v}_R - 2({\bf n}\cdot{\bf v}_R)^2 + 3v_R^2\right)n^i -
   (1-{\bf n}\cdot{\bf v}_R)v^i_R \nonumber\\
   &&- {2Gm\over R}{1\over\gamma_R^3(1-{\bf n}\cdot{\bf v}_R)}n^i
   \biggr] + \hbox{\it radiative terms}.
\label{b2}
\end{eqnarray}
   
As in the electromagnetic case (\ref{a7}), the leading nonradiative term in (\ref{b2}) is 
proportional to $n^i-v^i_R$, so to lowest order there is no aberration.  Now, however, there 
are additional corrections of higher order in $v$.  It is not hard to show that the effect of 
these corrections is to further ``extrapolate'' from the retarded position toward the 
``instantaneous'' position.  Indeed, one finds that
\beq
\Gamma^i_{00} = {Gm\over R^2}{1\over\gamma_R^2(1-{\bf n}\cdot{\bf v}_R)^2}
   \left[(1 + \epsilon_1)\eta^i  
   - {2Gm\over R}{1\over\gamma_R^4(1-{\bf n}\cdot{\bf v}_R)^4}n^i +
    \epsilon_2 v^i_R)\right] + \hbox{\it radiative terms} 
\label{b3}
\eeq
where
\beq
\eta^i = n^i + (t - z^0(s_R)){dn^i\over dt} + {1\over2}(t - z^0(s_R))^2{d^2n^i\over dt^2}
\label{b4}
\eeq
and $\epsilon_1$ and $\epsilon_2$ are of order $v^2$.  In other words, the gravitational 
acceleration is directed toward the retarded position of the source {\em quadratically\/} 
extrapolated toward its ``instantaneous'' position, up to small nonlinear terms and corrections 
of higher order in velocities.   

Does eqn.\ (\ref{b3}) imply that gravity propagates instantaneously?  As in the case of
electromagnetism, it clearly does not.  Every term in the connection $\Gamma^\rho_{\mu\nu}$
depends only on the retarded position, velocity, and acceleration of the source; despite Van 
Flandern's claim to the contrary \cite{TvF2}, there is no dependence, implicit or explicit,
on the ``instantaneous'' direction to the source.  Indeed, the vector (\ref{b4}) does not point 
toward the ``instantaneous'' position of the source, but only toward its position extrapolated 
from this retarded data.  In particular,  as in Maxwell's theory, if a source abruptly stops 
moving at a point $z(s_0)$, a test particle at position $x$ will continue to accelerate toward 
the extrapolated position of the source until the time it takes for a signal to propagate from 
$z(s_0)$ to $x$ at light speed.

A similar result can be obtained in general relativity by evaluating the gravitational field
of a boosted black hole \cite{Puthoff}, or more generally by systematically approximating 
the solution of the two-body problem \cite{Damourb}.  As in the case considered here, the 
gravitational interaction propagates at the speed of light, but velocity-dependent terms in
the interaction nearly cancel the effect of aberration.  Indeed, it can be rigorously proven 
that no gravitational influence in general relativity can travel faster than the speed of light 
\cite{Low}.

It is worth noting that the cancellation between aberration and velocity-dependent 
terms in general relativity is not quite exact.  If gravity could be described exactly as an 
instantaneous, central interaction, the mechanical energy and angular momentum of a 
system such as a binary pulsar would be exactly conserved, and orbits could not decay.  
In general relativity, the gravitational radiation reaction appears as a slight mismatch 
between the effects of aberration and the extra noncentral terms in the equations of motion 
\cite{Damourb}.  One could again try to formulate an alternative theory in which gravity 
propagated instantaneously, but, as in electromagnetism, only at the expense of ``deunifying'' 
the field equations and treating gravity and gravitational radiation as independent phenomena.

\section{Is the Cancellation a Miracle?}

We have seen that the observed lack of aberration in gravitational interactions need not 
imply an infinite propagation speed, but can be explained as the effect of velocity-dependent 
terms in the interaction.  There is still something to understand, though: a cancellation as 
exact as that of eqns.\ (\ref{a8}) and (\ref{b4}) must surely have a more fundamental origin.

A starting point is Lorentz invariance.  As Poincar{\'e} first observed, any Lorentz-invariant 
model of gravitation necessarily requires additional velocity-dependent interactions, 
which can provide ``a more or less perfect compensation'' for the effects of aberration 
\cite{Poin,Poin2}.  Indeed, Poincar{\'e} showed in Ref.\ \cite{Poin2} that for a 
Lorentz-invariant model of gravity with light-speed propagation, a correct Newtonian 
limit, and forces that depend only on positions and velocities, one can choose to eliminate 
all terms of order $v/c$, so that the deviations from Newtonian gravity are at most of order 
$v^2/c^2$.  Poincar{\'e} did not actually demonstrate that the cancellation of terms of 
order $v/c$ is {\em necessary\/},\footnote{In the discussion after sect.\ 9, eqn.\ (9) of Ref.\ 
\cite{Poin2}, Poincar{\'e}'s choice of setting the parameter $\beta$ to zero excluded terms 
of order $v/c$; a reintroduction of $\beta$ would restore such terms.} but he showed 
that aberration terms can be naturally excluded without doing violence to the theory. 

Consider, for example, the simplest Lorentz-invariant scalar theory of gravity.  The 
naive choice for a retarded Newtonian potential would be $\varphi = m/R$, where $R$ 
is the propagation-delayed distance (\ref{a4}).  The gradient $\nabla\varphi$ is proportional 
to $\bf n$, and exhibits aberration at order $v/c$.  But $\varphi$ is not Lorentz invariant;
the simplest invariant version is $\phi = m/r$, where $r$ is the retarded distance (\ref{a3}).  
It is easy to show that
\begin{eqnarray}
\nabla\phi &=&-{\gamma_R m\over r^2} {1\over(1-{\bf n}\cdot{\bf v}_R)}
   \left[(1-{\bf n}\cdot{\bf v}_R){\bf v}_R - (1-v_R^2){\bf n}\right] + \hbox{\it radiative terms}
   \\
   &=& {\gamma_R m\over r^2}\left( {\bf n} +  (t - z^0(s_R)){d{\bf n}\over dt}\right)
    + {\gamma_R m\over r^2}{1\over(1-{\bf n}\cdot{\bf v}_R)}
    {\bf v}_R\times({\bf v}_R\times{\bf n}) + \hbox{\it radiative terms} . \nonumber
\label{c1}
\end{eqnarray}
Thus up to terms of order $v^2/c^2$, the force points not toward the retarded position of 
the source, but toward the ``linearly extrapolated'' retarded direction (\ref{a8}); the extra 
velocity dependence in $r$ eliminates aberration at order $v/c$.

As Van Flandern has stressed, though, astronomical observations require a more complete 
cancellation:  aberration terms of order $v^3/c^3$ must be eliminated as well.  To 
understand such a cancellation, we can stand the argument of Ref.\ \cite{TvF} on its head.  
As that paper emphasized, a retarded purely central force with no velocity-dependent terms 
inevitably leads to the drastic nonconservation of orbital (``mechanical'') angular momentum 
and energy in a binary system.  But by Noether's theorem, any theory derived from a Lagrangian 
invariant under rotations and time translations must conserve total angular momentum and 
energy.  For an isolated, bound system, this is only possible if changes in mechanical angular 
momentum and energy are balanced by changes in the angular momentum and energy of 
radiation.

For electromagnetism, conservation of charge implies that there can be no monopole 
radiation, and the power radiated in dipole radiation is proportional to $|d^2{\bf d}/dt^2|^2$, 
where $\bf d$ is the electric dipole moment of the source.  Since the first derivative $d{\bf d}/dt$ 
is proportional to the velocity, a charge moving at a constant velocity can radiate no angular 
momentum or energy.  Hence at least to first order in velocity, any nonconservation of 
mechanical angular momentum and energy due to finite propagation speed {\em must\/} 
be compensated by additional (velocity-dependent) terms in the interaction.

To elaborate this argument, observe that by dimensional analysis, the radiated power is of the 
form $ P \sim |d^2{\bf d}/dt^2|^2 c^{-3} \sim e^2a^2/c^3$, where $a$ is the acceleration of
the source.  Moreover, the virial theorem implies that $mv^2 \sim e^2/r \sim mar$.  Thus
\beq
P \sim {e^2\over r^2}{v^4\over c^3} = F_{\hbox{\scriptsize Coulomb}}v\cdot{v^3\over c^3}
\label{c3}
\eeq
where $F_{\hbox{\scriptsize Coulomb}}$ is the lowest-order, ``instantaneous'' Coulomb 
force.  If we want this  radiated energy to balance the retardation-induced nonconservation of 
mechanical energy, the terms responsible for this nonconservation---the velocity-dependent 
corrections to $F_{\hbox{\scriptsize Coulomb}}$---must first appear at order $v^3/c^3$, 
and ``aberration'' effects must cancel at lower orders.  The exact mechanism for this cancellation 
may vary from theory to theory, but its existence is guaranteed by Noether's theorem, and we can 
be certain that it will appear in any field equations derived from an appropriately invariant action.

For gravity, conservation of momentum and angular momentum also rule out dipole radiation
\cite{MTW}.  The lowest order gravitational radiation is quadrupolar, and the radiated power 
goes as $|d^3{\bf Q}/dt^3|^2$, where $\bf Q$ is the mass quadrupole moment of the source.  A 
source with a constant second derivative of $\bf Q$ can therefore radiate no angular momentum 
or energy, and any nonconservation of mechanical angular momentum and energy must again 
be canceled by additional terms in the interaction.  The second derivative $d^2{\bf Q}/dt^2$ 
involves terms proportional to acceleration and to the square of the velocity, so the cancellation 
must occur at a higher order than it did for electromagnetism.  Dimensional analysis now yields
\beq
P \sim F_{\hbox{\scriptsize Newton}}v\cdot{v^5\over c^5} ,
\label{c4}
\eeq
so this cancellation must be present up to order $v^3/c^3$, in agreement with observation. 

Note that the dipolar nature of electromagnetic radiation is intimately tied to the fact that 
the electromagnetic interaction is vectorial (spin 1) \cite{Couch}.  Similarly, the quadrupolar 
nature of gravitational radiation is tied to the traceless tensorial (spin 2) form of the interaction.  
The case of a scalar interaction is a bit more subtle.  For a non-Lorentz invariant theory, 
conservation laws place no restrictions on radiation, and monopole radiation should dominate, 
permitting aberration at all orders.  For a Lorentz-invariant theory, though, the standard 
coupling of a scalar field to matter dictates that the monopole moment has the form $m = 
m_0F(\phi)$, so ${\dot m}\sim{\dot\phi}$, which is suppressed by the equations of motion 
\cite{Farese}.  The leading contribution to energy loss thus comes from dipole radiation, and 
as with electrodynamics, aberration terms first appear at order $v^3/c^3$.  This is the case 
for the interaction (\ref{c1}).\footnote{Note that terms of even order in $v/c$ are invariant 
under time reversal, and do not lead to secular changes, though strictly speaking one must 
check that there are no time-reversal-invariant effects with very long periods, which can 
mimic secular changes.}

Finally, let us return to the question asked in Ref.\ \cite{TvF}: what do experiments say about 
the speed of gravity?  The answer, unfortunately, is that so far they say fairly little.  In the absence 
of direct measurements of propagation speed, observations must be filtered through theory, and 
different theoretical assumptions lead to different deductions.  In particular, while the observed 
absence of aberration is consistent with instantaneous propagation (with an extra interaction 
somehow added on to explain the gravitational radiation reaction), it is also consistent with the 
speed-of-light propagation predicted by general relativity.

Within the framework of general relativity, though, observations do give an answer.  The 
Einstein field equations contain a single parameter $c_g$, which describes both the speed of 
gravitational waves and the ``speed of gravity'' occurring in the expression for aberration and 
in the velocity-dependent terms in the interaction.  This parameter appears in the gravitational 
radiation reaction in the form $c_g^{-5}$, as in eqn.\ (\ref{c4}), and the success of the theory 
in explaining the orbital decay of binary pulsars implies that $c_g=c$ at the $1\%$ level or 
better \cite{Taylor}.

\vspace{1.5ex}
\begin{flushleft}
\large\bf Acknowledgements
\end{flushleft}

This work was supported in part by Department of Energy grant DE-FG03-91ER40674.

\end{document}